\documentclass[preprint,onecolumn,nofootinbib]{revtex4}
\pdfoutput=1
\usepackage[colorlinks=true,linkcolor=blue,urlcolor=blue,filecolor=black,citecolor=red,
pdfstartview=FitV,pdftitle={},pdfsubject={},pdfkeywords={},pdfpagemode=None,bookmarksopen=true]{hyperref}
\usepackage{graphicx}%Include figure files
\usepackage{amsmath}
\usepackage{amsfonts}
\usepackage{amssymb,ulem}
\usepackage{color}%
\usepackage{CJK}
\usepackage{dcolumn}% Align table columns on decimal pointl
\newcommand{\f}{\begin{equation}}
\newcommand{\ff}{\end{equation}}
\newcommand{\fa}{\begin{eqnarray}}
\newcommand{\ffa}{\end{eqnarray}}

\begin{document}

\title{The Effect of Anisotropy on Holographic Entanglement Entropy and Mutual Information}
\author{Peng Liu $^{1}$}
\email{phylp@jnu.edu.cn}
\author{Chao Niu $^{1}$}
\email{niuchaophy@gmail.com}
\author{Jian-Pin Wu $^{2,3}$}
\email{jianpinwu@yzu.edu.cn}
\affiliation{
$^{1}$ Department of Physics and Siyuan Laboratory, Jinan University, Guangzhou 510632, P.R. China\\
$^{2}$ Center for Gravitation and Cosmology, College of Physical Science and Technology, Yangzhou University, Yangzhou 225009, China\\
$^{3}$ School of Aeronautics and Astronautics, Shanghai Jiao Tong University, Shanghai 200240, China
}
\begin{abstract}
We study the effect of anisotropy on holographic entanglement entropy (HEE) and holographic mutual information (MI) in the Q-lattice model, by exploring the HEE and MI for infinite strips along arbitrary directions. We find that the lattice always enhances the HEE. The MI, however, is enhanced by lattice for large sub-regions; while for small sub-regions, the MI is suppressed by the lattice. We also discuss how these phenomena result from the deformation of geometry caused by Q-lattices.

\end{abstract}
\maketitle
\tableofcontents
\section{Introduction}

Anisotropy is universal and results in many rich phenomena in nature, such as magnetic systems, latticed systems, and so on \cite{Glotzer:2007}. In some strongly correlated systems, the anisotropy is associated with many entanglement measures, and have novel applications in measuring instruments. For example, the quantum entanglement can be exploited to design magnetic compass sensors \cite{cai:2010,gauger:2011,hannah:2012}. Moreover, the entanglement can associate with physical observables for anisotropic quantum phase transitions \cite{Somma:2004}. The effects of anisotropy on entanglement structure for strongly correlated systems are useful for practical uses and are worthy of further investigation. Strongly correlated systems, however, are long-standing hard problems in physics; entanglement is also hard to study. Gauge/gravity duality can bring together the strongly correlated systems and entanglement and offer a good platform to study the anisotropy of entanglement in strongly correlated systems.

Gauge/gravity duality has been proved powerful tools to study strongly correlated systems and quantum information properties \cite{Maldacena:1997,Witten:1998,Ryu:2006bv}.  Anisotropy is also ubiquitous in holographic systems, such as systems with lattices, anisotropic axions, massive gravity and so on \cite{Ling:2015ghh,Fang:2014jka,Arefeva:2018hyo}. All these models realize the anisotropy by explicitly breaking the isotropic symmetry. The anisotropy can also be introduced by spontaneous symmetry breaking, such as holographic charge density wave models \cite{Donos:2013gda,Ling:2014saa}. Especially, the latticed structure plays a crucial role in obtaining finite direct current transportation coefficients, Mott insulator, metal-insulator transitions \cite{Donos:2013eha,Ling:2015exa}.

Another huge advantage of gauge/gravity duality is the amazing connection between information-related quantities and geometrical quantities. The entanglement entropy (EE), a commonly accepted entanglement measure, was proposed to be proportional to the minimum surface area. This geometric prescription, referred to as holographic entanglement entropy (HEE), has been extensively studied and applied in the study of phase transitions, etc
\cite{Nishioka:2006gr,Klebanov:2007ws,Pakman:2008ui,Fujita:2009kw,Kuang:2014kha,Ling:2015dma,Ling:2016wyr,Ling:2016dck,Ling:2017naw,Zeng:2016fsb,Baggioli:2018afg,Zhang:2016rcm}. Besides that, many new information-related quantities have been proposed to have geometrical duals. The mutual information (MI), whose definition derives from the HEE, reveals more details of entanglement structures of quantum systems. Also, the R\'enyi entropy has been proposed as proportional to the minimal area of the cosmic brane \cite{Dong:2016fnf}. The entanglement of purification, which involves the purification of the mixed states, have been associated with the area of the minimal cross section of the entanglement wedge \cite{Takayanagi:2017knl}. Information related quantities are becoming the core of the holographic theories.

Despite its power in measuring the pure state entanglement, EE is unsuitable for measuring the mixed state entanglement. Many other measures have been proposed to measure the mixed state entanglement, such as mutual information, entanglement of purification, non-negativity, and so on - all these have holographic duals \cite{Chaturvedi:2016rft,Chaturvedi:2016rcn,Takayanagi:2017knl}. The geometric interpretations of entanglement measures greatly simplify the study of entanglement structures in strongly correlated systems.

We study the effect of anisotropy on HEE and MI (for the infinite strip) in the holographic Q-lattice model. We find that the Q-lattice always enhances the HEE, regardless of the system parameters and the size of the subregion. The Q-lattice effects on MI, however, depend on the size of the subregions: for small subregions the Q-lattice enhance the MI; when the subregion enlarges the Q-lattice effect become non-monotonic; for large enough subregions the Q-lattice suppresses the MI. We also discuss how these phenomena result from the deformation of the geometry caused by Q-lattices. These results deepen our understanding of how the anisotropy affects entanglement measures and can stimulate further investigations on this new topic. Previous studies on anisotropic effects on entanglement related quantities can be found in \cite{Ahn:2017kvc,Jahnke:2017iwi,Avila:2018sqf,Jokela:2019ebz,Jokela:2019tsb,Dudal:2018ztm,Dey:2014voa,Mishra:2016yor,Mishra:2018tzj,Gursoy:2018ydr,Roychowdhury:2015fxf,Mahapatra:2019uql,Narayan:2012ks,Narayan:2013qga,Mukherjee:2014gia,Narayan:2015lka,Giataganas:2012zy,Giataganas:2013lga}.

We organize this paper as three parts: we review the Q-lattice model and deduce the anisotropic HEE in \ref{qla-ani}; then we study the anisotropic HEE and MI in \ref{subsec:heearbi}; we conclude and discuss in \ref{sec:discussion}.

\section{Holographic Q-lattice model and Anisotropic Holographic Entanglement Entropy}\label{qla-ani}
\subsection{Holographic Q-lattice model}
The holographic Q-lattice model is a concise realization of the periodic structure. Previous holographic lattice models, such as the ionic lattices model and the scalar lattices model, introduces spatially periodic structures on scalar fields or the chemical potential (see \cite{Ling:2015ghh} for a recent review). The resultant equations of motion are a set of highly nonlinear partial differential equations, which poses a challenge for numerical solutions. By contrast, the Q-lattice model introduces a complex scalar field, which results in only ordinary differential equations. Therefore, the Q-lattice model is an easier realization of lattice structures. The Q-lattice model is useful in modeling the Mott insulator and metal-insulator transitions \cite{Ling:2015exa,Ling:2015epa}.

The Lagrangian of the Q-lattice model is \cite{Donos:2013eha,Donos:2014uba,Ling14laa,Ling:2015dma},
\begin{equation}\label{actionq}
\mathcal L = R + 6 - \frac{F^{ab} F_{ab}}{2} - \partial_a \Phi^\dagger \partial^a \Phi -  m^2 |\Phi|^2.
\end{equation}
System \eqref{actionq} can be solved with ansatz,
\begin{equation}\label{ansatz}
	\begin{aligned}
	d s ^ { 2 } & = \frac { 1 } { z ^ { 2 } } \left[ - ( 1 - z ) p ( z ) U d t ^ { 2 } + \frac { d z ^ { 2 } } { ( 1 - z ) p ( z ) U } + V _ { 1 } d x ^ { 2 } + V _ { 2 } d y ^ { 2 } \right], \\
	A & = \mu ( 1 - z ) a d t, \qquad \Phi  = e ^ { i \tilde k x } z ^ { 3 - \Delta } \phi,
	\end{aligned}
\end{equation}
where $p ( z ) = 1 + z + z ^ { 2 } - \mu ^ { 2 } z ^ { 3 } / 2 $ and $ \Delta = 3 / 2 \pm \left( 9 / 4 + m ^ { 2 } \right) ^ { 1 / 2 }$. We set $m^{2} = -2$ for concreteness, and hence $\Delta = 2$. The horizon and the boundary locate at $z=1$ and $z=0$ respectively. The $A_{a}$ is the Maxwell field, and $\phi$ is the complex scalar field mimicking the lattice structures. Consequently, the functions to solve are $\left(a,\phi,U,V_{1},V_{2}\right)$.

In order to solve the system \eqref{actionq}, we need to specify the boundary conditions and system parameters. We set $a(0)=1$, then $A_{t}(0) = \mu$ becomes the chemical potential of the dual system. The boundary condition $\tilde\lambda= \phi(0)$ is the strength of the lattice deformation, and $\tilde k$ is the wave vector of the periodic structure. The asymptotic AdS$_{4}$ requires that $U(0)=1, V_{1}(0)=1,V_{2}(0)=1$. Other boundary conditions at the horizon $(z=1)$ can be fixed by regularity. The Hawking temperature reads $\tilde T = \left( 6 - \mu ^ { 2 } \right) U ( 1 ) / ( 8 \pi )$. The black brane solutions can be categorized by $3$ dimensionless parameters $\left(T,\lambda,k\right)\equiv\left(\frac{\tilde T}{\mu},\frac{\tilde \lambda}{\mu},\frac{\tilde k}{\mu}\right)$, where we adopt the chemical potential $\mu$ as the scaling unit.

\subsection{Anisotropic Holographic entanglement: HEE over arbitrary direction}\label{subsec:heearbi}

The HEE of subregion $A$ is
\begin{equation}\label{def:hee}
S _ { A } = \frac { \operatorname { Area } \left( \Sigma _ { A } \right) } { 4 G _ { N } ^ { ( d + 2 ) } }.
\end{equation}
where $\Sigma_{A}$ is the minimal surface satisfying $\partial A = \partial\Sigma_{A}$ \cite{Ryu:2006bv}. The HEE of many subregions, such as disks, infinite strips and cusps, has been widely studied in holographic systems. Apparently, only HEE of non-circular subregions is sensitive to the anisotropy.

\begin{figure}[htbp]
	\centering
	\includegraphics[width=0.5\textwidth]{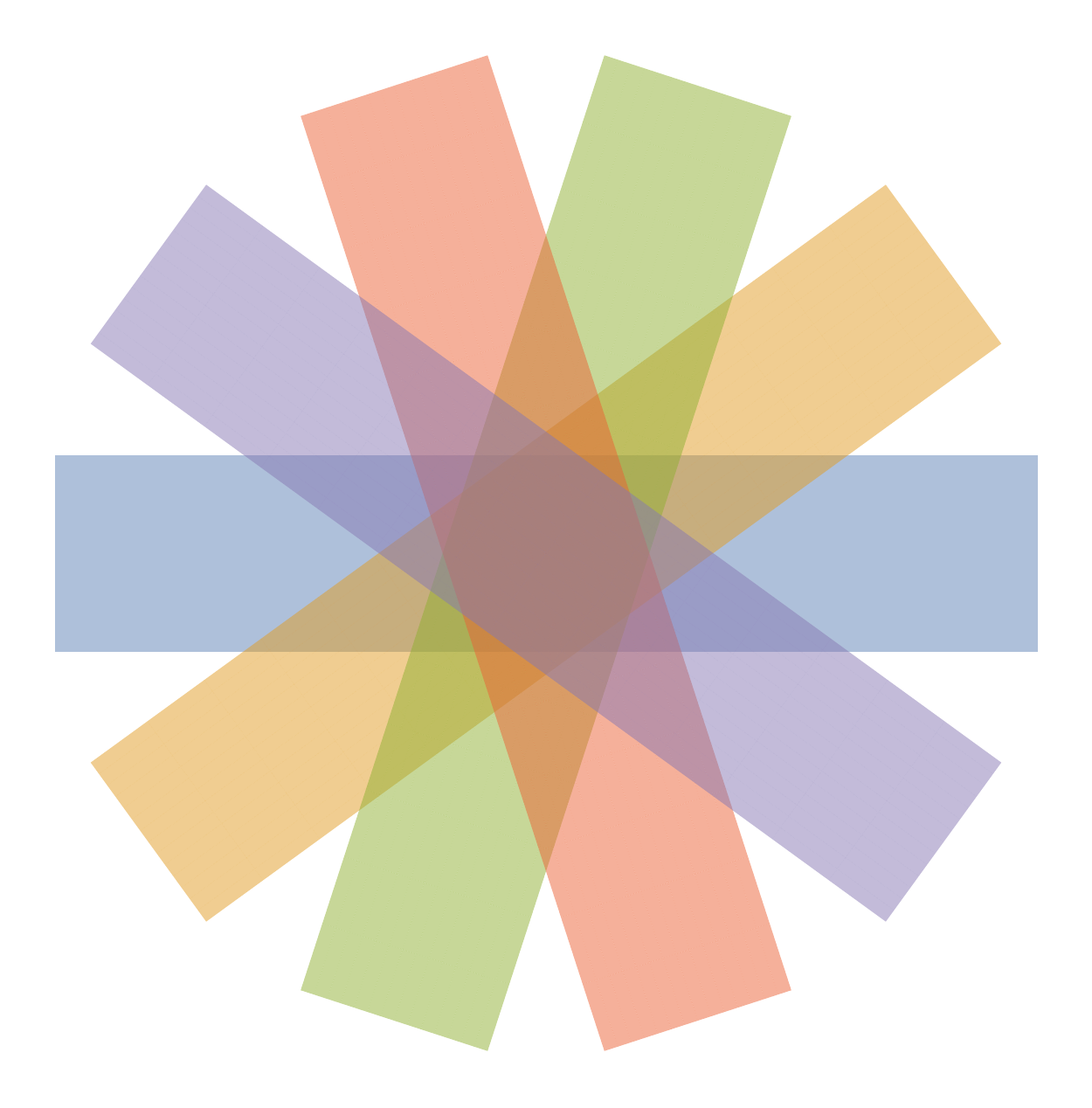}
	\caption{The demonstration of strips pointing at different directions.}
	\label{fig:rotate}
\end{figure}

For simplicity, we consider the HEE of infinite strip partition for $4$-dimensional homogeneous geometry\footnote{Our deduction can also be applied to systems with off-diagonal metric.},
\begin{equation}\label{geo:arbitrary}
	ds^{2} = g_{tt} dt^{2} + g_{zz} dz^{2} + g_{xx} dx^{2} + g_{yy} dy^{2}.
\end{equation}
Previous researches on HEE usually adopt the infinite strip along a fixed direction at which the HEE does not capture the anisotropy effect.
It is important to study how anisotropy affect the entanglement structure. To this end, we study the HEE of infinite strips along arbitrary directions (see Fig. \ref{fig:rotate}), which we call as anisotropy HEE.

For an infinite strip pointing at direction $\vec \theta = \left(\cos \theta, \sin \theta\right)$ on $(x,y)$-plane, the minimal surface will be invariant along $\vec \theta$. It is then more convenient to work in a new coordinate
\begin{equation}\label{newcord}
	\tilde t = t, \quad \tilde z = z,\quad \tilde{x} = x \cos (\theta )+y \sin (\theta ),\quad \tilde y = y \cos (\theta )-x \sin (\theta ),
\end{equation}
where the direction $\vec\theta$ is along $\tilde x$ in coordinate system \eqref{newcord}. The minimal surface is invariant along $\tilde x$, and hence the minimal surface can be described by $\tilde z\left(\tilde y\right)$. In coordinate system \eqref{newcord} the geometry \eqref{geo:arbitrary} is written as,
\begin{equation}
	\begin{aligned}
	ds^{2} = &g_{\tilde t\tilde t} d\tilde t^{2} + g_{\tilde z\tilde z} d\tilde z^{2} - \left(g_{xx}-g_{yy}\right)\sin (2\theta )d\tilde x d \tilde y\\
	& + \left(g_{xx} \cos ^2(\theta )+g_{yy} \sin ^2(\theta )\right)d\tilde x^2 + \left(g_{xx} \sin ^2(\theta )+g_{yy} \cos ^2(\theta )\right)d\tilde y^2.
	\end{aligned}
	\label{newgeo}
\end{equation}
The induced metric on the hypersurface $\tilde z(\tilde y)$ at $\tilde t = \text{const}$ reads,
\begin{equation}
	\begin{aligned}
	d\hat s^{2}  & = g_{\tilde x \tilde x} d\tilde x^2 +  g_{\tilde z\tilde z} d\tilde z^{2} + g_{\tilde y \tilde y} d\tilde y^2 +  g_{\tilde x \tilde y} d\tilde x d\tilde y \\
	& = g_{\tilde x \tilde x} d\tilde x^2 + \left(g_{zz} z'\left(\tilde y\right)^{2}  + g_{\tilde y\tilde y} \right) d \tilde y^{2} + g_{\tilde x \tilde y} d\tilde x d\tilde y.
	\end{aligned}
	\label{inducedmetric}
\end{equation}
Therefore the area of the minimal surface is
\begin{equation}
	A = \int_{\Sigma} \mathcal L d\tilde x d\tilde y = L_{x}\int_{\Sigma} \mathcal L d\tilde y.
	\label{minimalarea}
\end{equation}
where $\mathcal L = \sqrt{\hat g} = \sqrt{g_{{xx}} g_{{yy}}+ g_{{zz}} z'({\tilde y})^2 \left(g_{{xx}} \cos ^2(\theta )+g_{{yy}} \sin ^2(\theta )\right) }$, and $L_{x} \equiv \int d\tilde x$ is the length of the infinite strip along $\tilde x$.
For simplicity we ignore some common factors and denote the HEE as,
\begin{equation}\label{def:hee_om}
\hat S_{0} \equiv \int_{\Sigma} \mathcal L d\tilde y.
\end{equation}
The integration \eqref{def:hee_om} diverges as $\frac{2}{\epsilon}$ with $\epsilon$ the cut-off, due to the asymptotic AdS$_{4}$ boundary. One can extract the finite part of the HEE by subtracting a common divergence $\frac{2}{\epsilon}$,
\begin{equation}\label{eq:def_hee_finite}
\hat S = \hat S_{0} - \frac{2}{\epsilon}.
\end{equation}
Adopting the $\mu$ as the scaling unit, the scale invariant width and HEE is given by $l\equiv \hat l \mu$ and $S\equiv \hat S/\mu$ with $\hat l = \int d\tilde y$.

Treating the \eqref{def:hee_om} as a Lagrangian independent of $\tilde y$, the corresponding Hamiltonian is a constant along the minimal surface $\tilde z (\tilde y)$. The homogeneity of the background requires that a minimal surface shall reach a local bottom at some $\tilde z_{*}$, with which the width and $l$ the HEE $S$ can be uniquely decided $l(\tilde z_{*})$ and $S(\tilde z_{*})$.

Given the algorithm to compute the $l$ and $S$ we turn to study the anisotropy effects on HEE and MI.

\section{Anisotropic Holographic Entanglement Entropy in Q-lattice model}
The HEE is dictated by the background geometry, which spans over $(\lambda,k,T)$ for the Q-lattice model. The $(\lambda, k)$ reflects the strength of the lattice and the wavelength of the lattice respectively. For $\lambda =0$ the lattice is absent, the system reduces to AdS-RN black hole; while for $k=0$, the translational invariance and isotropy are recovered. The anisotropy effect is therefore only significant for sufficiently large values of $\lambda$ and $k$.

\begin{figure}[htbp]
	\centering
	\includegraphics[width=0.45\textwidth]{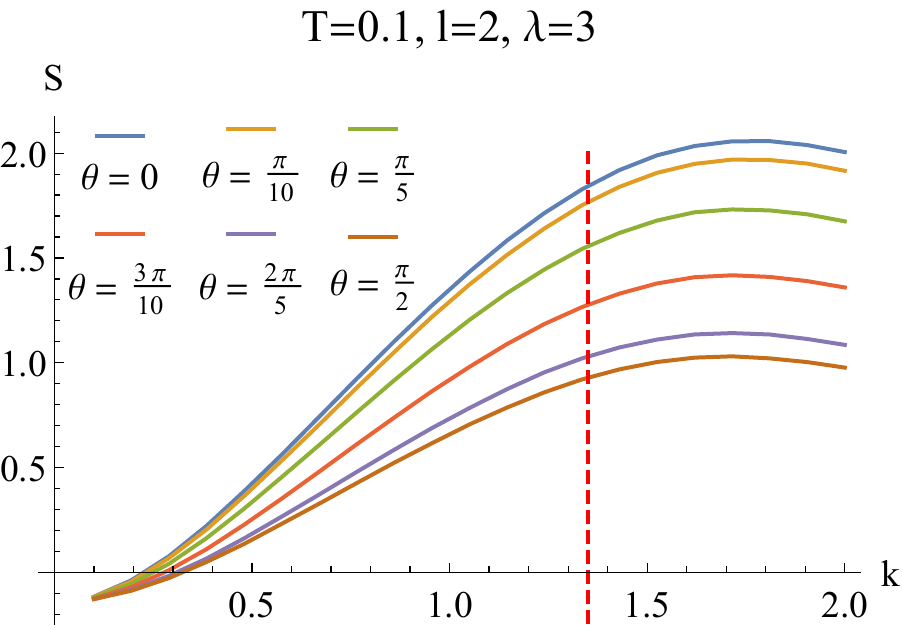}
	\includegraphics[width=0.45\textwidth]{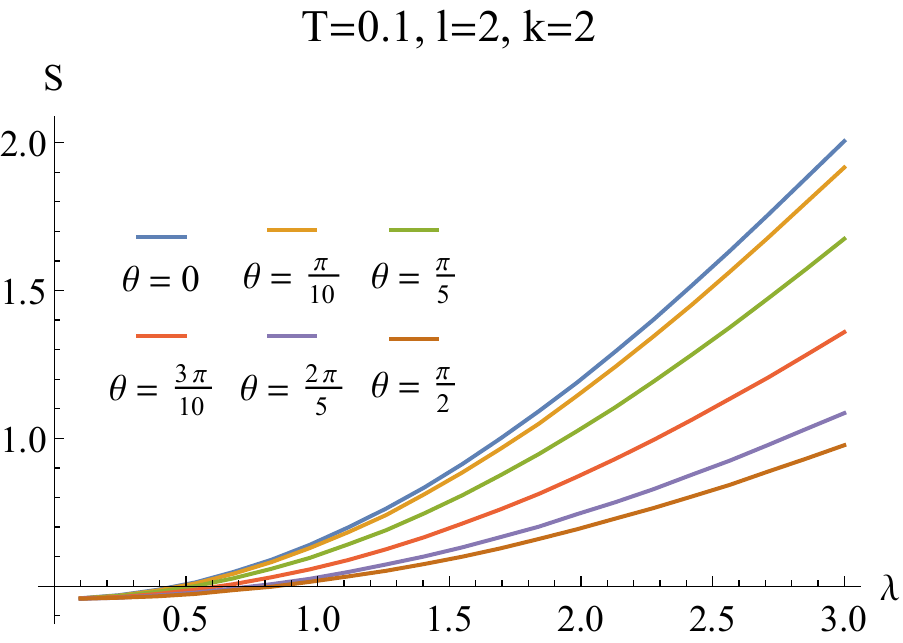}
	\caption{The left (right) plot is the HEE at different angle as function of $k\,(\lambda)$ respectively. The red vertical line in the left plot is the critical point of the metal-insulator transition.}
	\label{fig:lambdakdir}
\end{figure}

First, we reveal the entanglement structure by studying the relation between the HEE and parameter $\left(\lambda,k,\theta\right)$, at a typical temperature $T=0.1$\footnote{Similar phenomena are found in low-temperature regions.}. As depicted in Fig. \ref{fig:lambdakdir}, the HEE in arbitrary direction is all close to each other at small values of $\lambda$ or $k$.\footnote{From the parity of \eqref{minimalarea} we see that only $\theta\in [0,\pi/2]$ is needed to be explored.}. This is the reflection of the fact that the lattice effect is only significant for large enough values of $\lambda$ and $k$.

Moreover, the left plot of Fig. \ref{fig:lambdakdir} suggests that the HEE exhibits some extremal behavior near the quantum critical points of the metal-insulator transition \cite{Ling:2015dma}. This phenomenon shows that metal-insulator transitions can be characterized by the HEE in an arbitrary direction, not just by the HEE in the direction perpendicular to the lattice (see \cite{Ling:2015dma}). This phenomenon is expected because the HEE is largely supported by thermal entropy for relatively large subregions, whose HEE reads,
\begin{equation}\label{eq:dire-lag}
S =\int_{\Sigma} \mathcal L d\tilde y=\int_{\Sigma}  \sqrt{g_{{xx}} g_{{yy}}+ g_{{zz}} z'({\tilde y})^2 \left(g_{{xx}} \cos ^2(\theta )+g_{{yy}} \sin ^2(\theta )\right) } d\tilde y \simeq \int_{\Sigma} \sqrt{g_{xx} g_{yy}} d\tilde y.
\end{equation}
Therefore, the fact that thermal entropy characterizes the metal-insulator transitions results in the phenomenon that HEE along arbitrary direction characterizes the metal-insulator transitions. We also remark that the relation $S$ vs $\lambda$ can also exhibits extremal behavior near the quantum critical points, as long as the temperature is low enough.

\begin{figure}[htbp]
	\centering
	\includegraphics[width=0.49\textwidth]{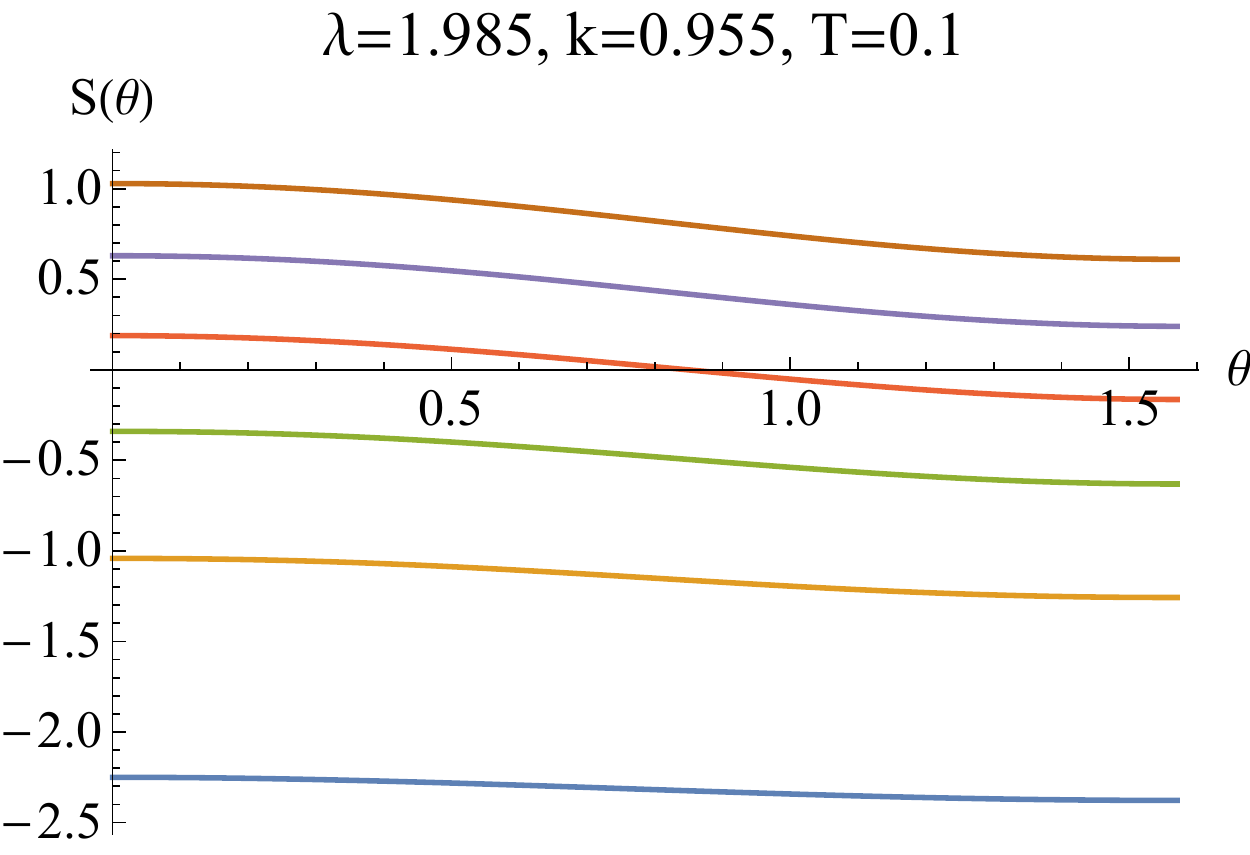}
	\includegraphics[width=0.49\textwidth]{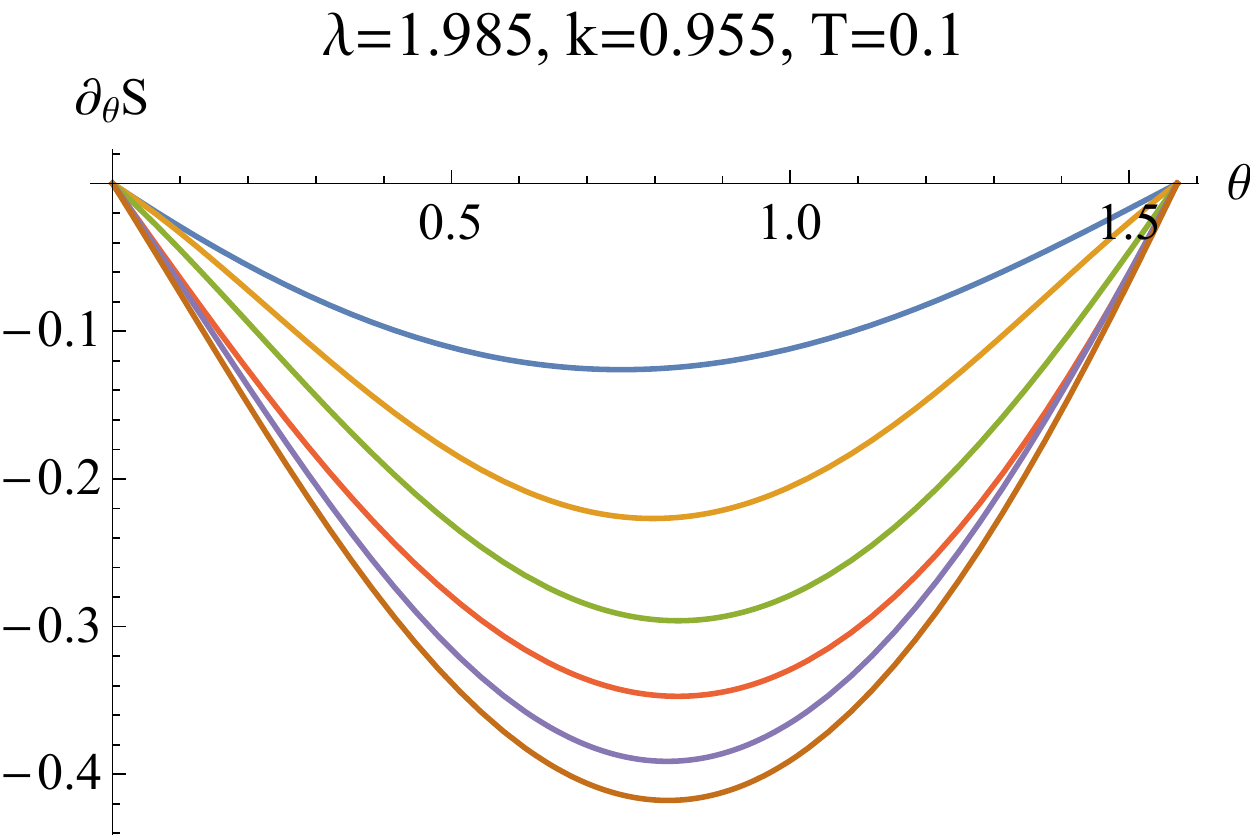}
	\includegraphics[width=0.49\textwidth]{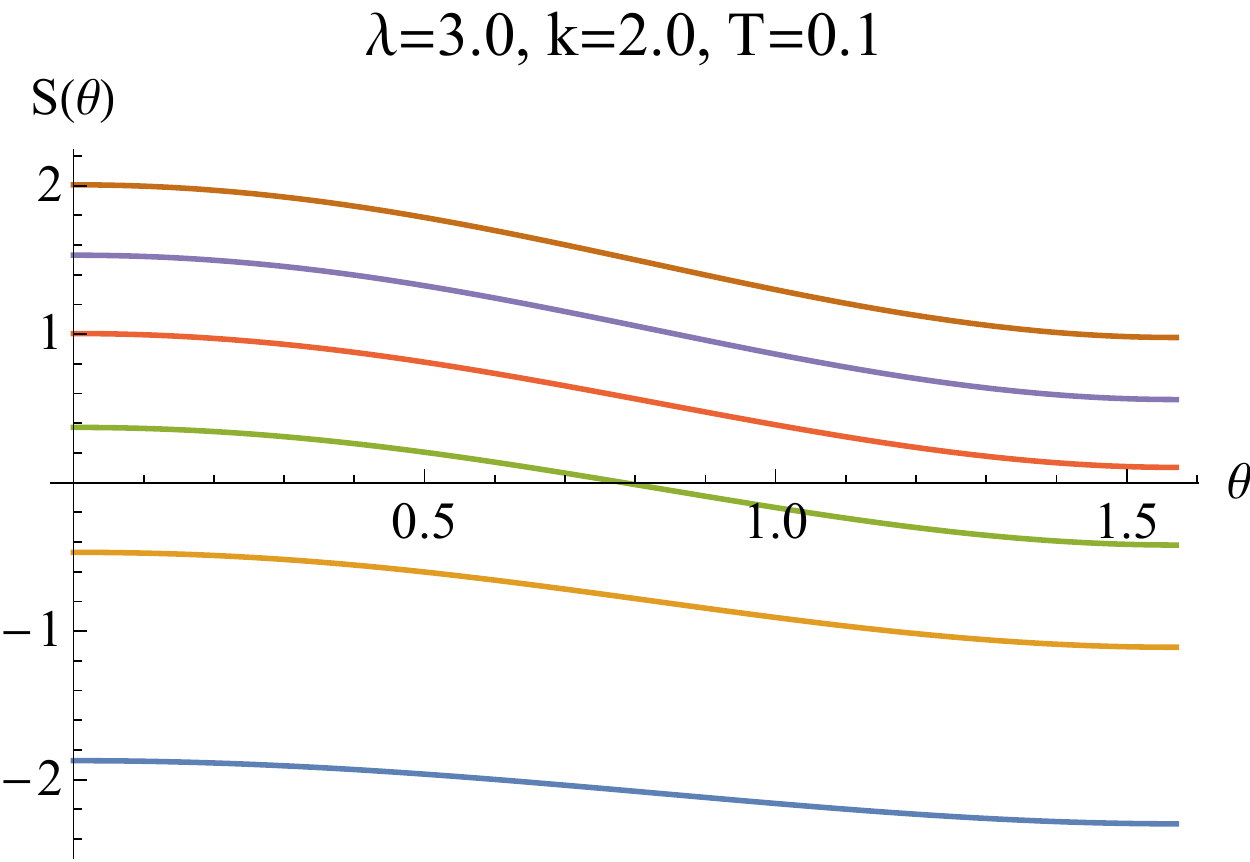}
	\includegraphics[width=0.49\textwidth]{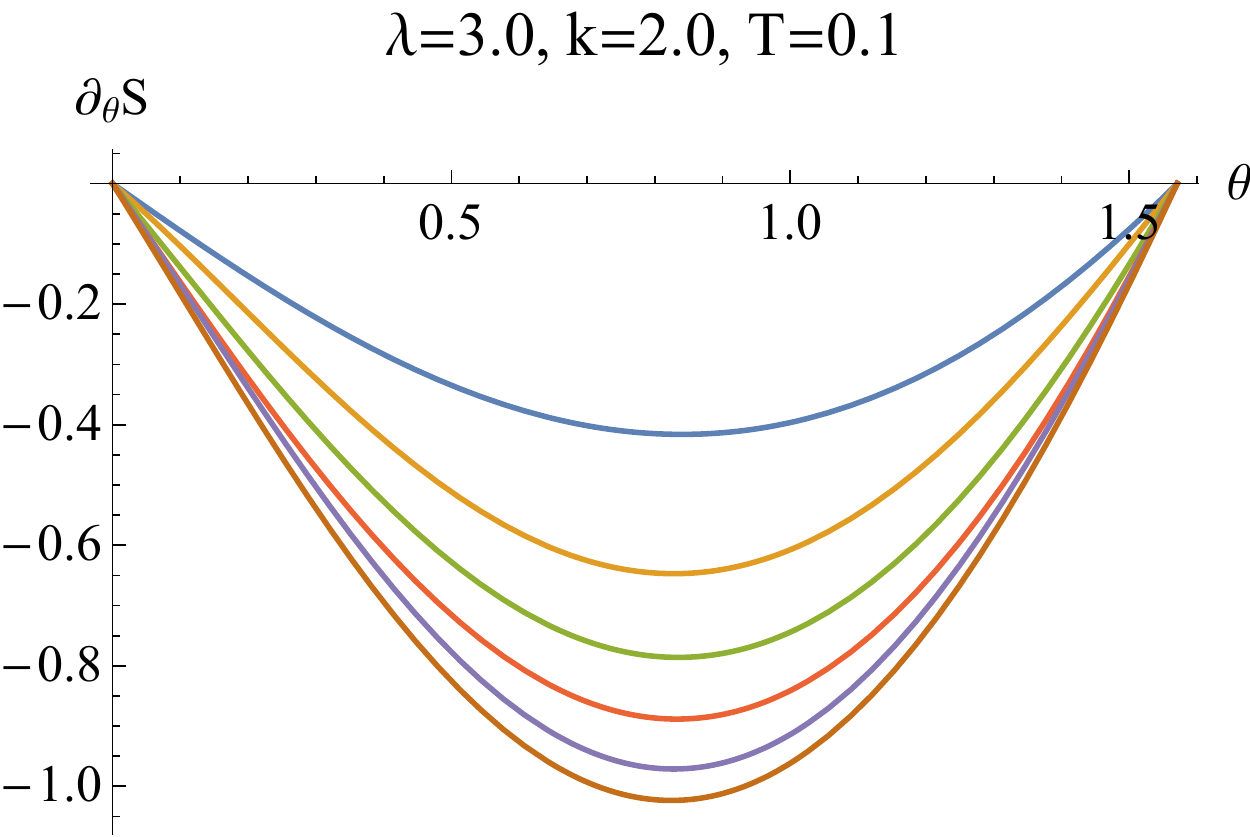}
	\includegraphics[width=0.65\textwidth]{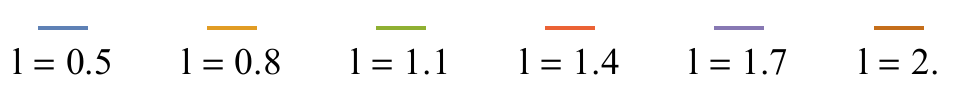}
	\caption{
	First row: $S$ vs $\theta$ (left plot) and $\partial_{\theta} S$ (right plot) for $(\lambda,k) = (1.985,0.955)$. 
	Second row: $S$ vs $\theta$ (left plot) and $\partial_{\theta} S$ (right plot) for $(\lambda,k) = (3,2)$.  
	Each curve in above four plots corresponds to different values of $l$ marked by the legends below.}
	\label{fig:overlambdak}
\end{figure}

Fig. \ref{fig:lambdakdir} also shows that the HEE decreases with $\theta$, regardless of the values of $\lambda,k$ and $T$. This phenomenon indicates that the lattice always enhances the HEE, noticing that the lattice points at $x$-direction (corresponds to $\theta=0$). To demonstrate more comprehensive content of anisotropic effect of HEE, we show the angular dependence of HEE in Fig. \ref{fig:overlambdak}. The first and the second row in Fig. \ref{fig:overlambdak} are $S$ vs $\theta$ (left plot) and $\partial_{\theta} S$ (right plot) for $(\lambda,k) = (1.985,0.955)$ and $(3,2)$, respectively. Comparing plots from the first row with those from the second row, it can be seen that when the $(\lambda,k)$ is larger, $S$ changes faster with $\theta$. That is, the anisotropy is more obvious for larger values of $(\lambda,k)$. This phenomenon is as expected - the anisotropy is more pronounced when the lattice effect becomes stronger. We also see from Fig. \ref{fig:overlambdak} that the $\partial_{\theta} S|_{\theta=0,\pi/2} =0$, which is a simple reflection of the fact that $\partial_{\theta} \mathcal L \sim \sin\left(2\theta\right)$.

\begin{figure}[htbp]
	\centering
	\includegraphics[width=0.55\textwidth]{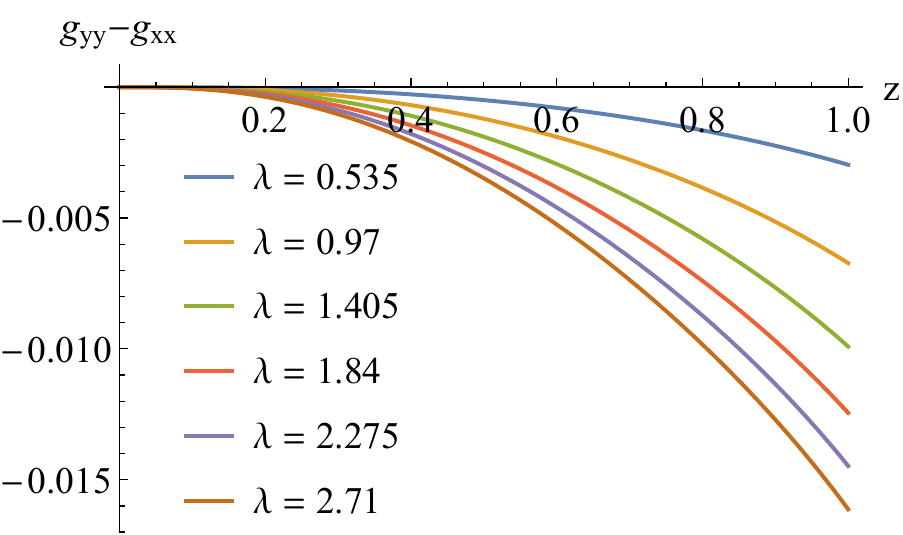}
	\caption{$g_{yy}-g_{xx}$ are all negative at $k=0.1, T=0.1$.
	}
	\label{fig:gyymgxx}
\end{figure}

Another important feature of the anisotropy effect on HEE is that the monotonic behavior of $S(\theta)$ is independent of values of $l$.
This feature could be understood from the geometry deformed by Q-lattices. The monotonically decreasing behavior of $S(\theta)$ means that

	\begin{equation}\label{eq:dthetaL}
	\partial_{\theta} \mathcal L = \frac{g_{zz} \sin\left(2\theta\right) \left(g_{yy}-g_{xx}\right) z'(x)^2}{2\sqrt{g_{zz} z'(x)^2 \left(g_{xx} \cos ^2(\theta )+g_{yy} \sin ^2(\theta )\right)+g_{xx} g_{yy}}} < 0.
	\end{equation}
Eq. \eqref{eq:dthetaL} suggests that $g_{yy}-g_{xx}<0$ for the angular range we consider.
We find $g_{yy}-g_{xx}<0$ for all values of $(\lambda,k,T)$ indeed (see Fig. \ref{fig:gyymgxx}). Therefore it is the geometry deformed by the Q-lattice that leads to the phenomena that Q-lattice always enhance the HEE. Notice also that $\left(g_{yy}-g_{xx}\right)|_{z\to 0}\to 0$ is due to the boundary condition $g_{xx}(0) = g_{yy}(0) = 1$. The monotonic behavior of $S(\theta)$ is apparently model-dependent, the scenario could be more diverse for other holographic models.

The HEE is a good measure of pure state entanglement, while not appropriate for characterizing mixed state entanglement. Especially, the thermal entropy for large subregions starts to contribute to the HEE \cite{Fischler:2012uv} and subordinate the quantum entanglement. We study the MI structure over the anisotropic Q-lattice model in order to further understand the entanglement.

\section{Anisotropic Mutual Information in Q-lattice model}

\begin{figure}[htbp]
	\centering
	\includegraphics[width=0.55\textwidth]{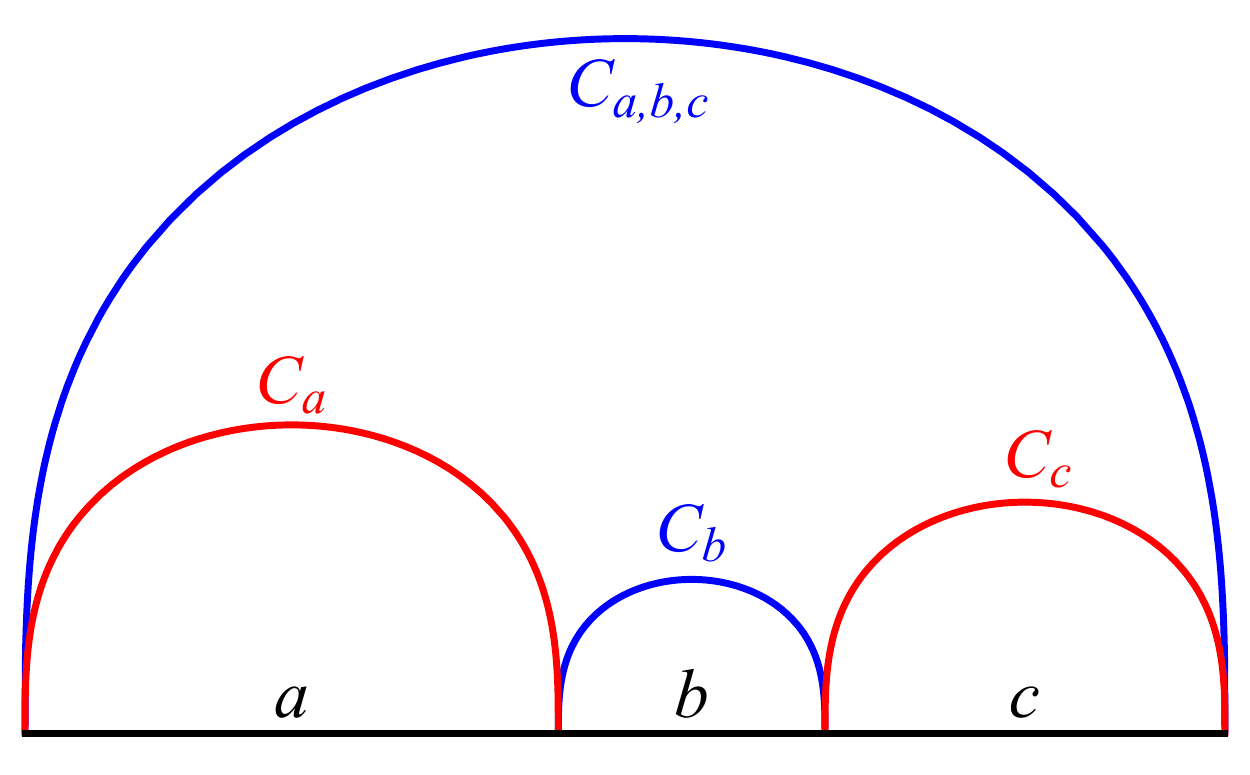}
	\caption{MI demonstration. The black line is the AdS boundary, and $a,c$ are the widths of the infinite strips $A$ and $C$, and $b$ is the width of the separation $B$. The $C_{a},C_{b},C_{c}$ and $C_{a,b,c}$ represents the minimal curve ending on $a,b,c$ and $a+b+c$.}
	\label{fig:mideo}
\end{figure}

The mutual information $I(A;C)$ measures the entanglement between two separate subregions $A$ and $C$.
\begin{equation}\label{midef}
I(A;C):=S(A)+S(C)-S(A\cup C).
\end{equation}
There are two configurations of $S(A\cup C)$ with locally minimal area, the blue ones and red ones (see Fig. \ref{fig:mideo} for demonstration). The definition of HEE requires the global minimum, {\it i.e.}, $S(A\cup C)=\min \{ S(A) + S(C), S(B) + S(A\cup B\cup C) \}$. Therefore the MI is,
\begin{equation}\label{mi:value}
I(A;C) = \left\{
\begin{aligned}
& 0, \\
& S(A)+S(C) - S(B) - S(A\cup B\cup C).
\end{aligned}
\right.
\end{equation}
The definition of MI \eqref{midef} not only cancels out the area law divergence, but also the volume law from the thermal contribution \cite{Fischler:2012uv}.
We study the MI structure of the parallel infinite stripes. Given a two-party system with $A\cup C$ with the separation $B$, we demonstrate the MI structure over the angle $\theta$.

\begin{figure}[htbp]
	\centering
	\includegraphics[width=0.49\textwidth]{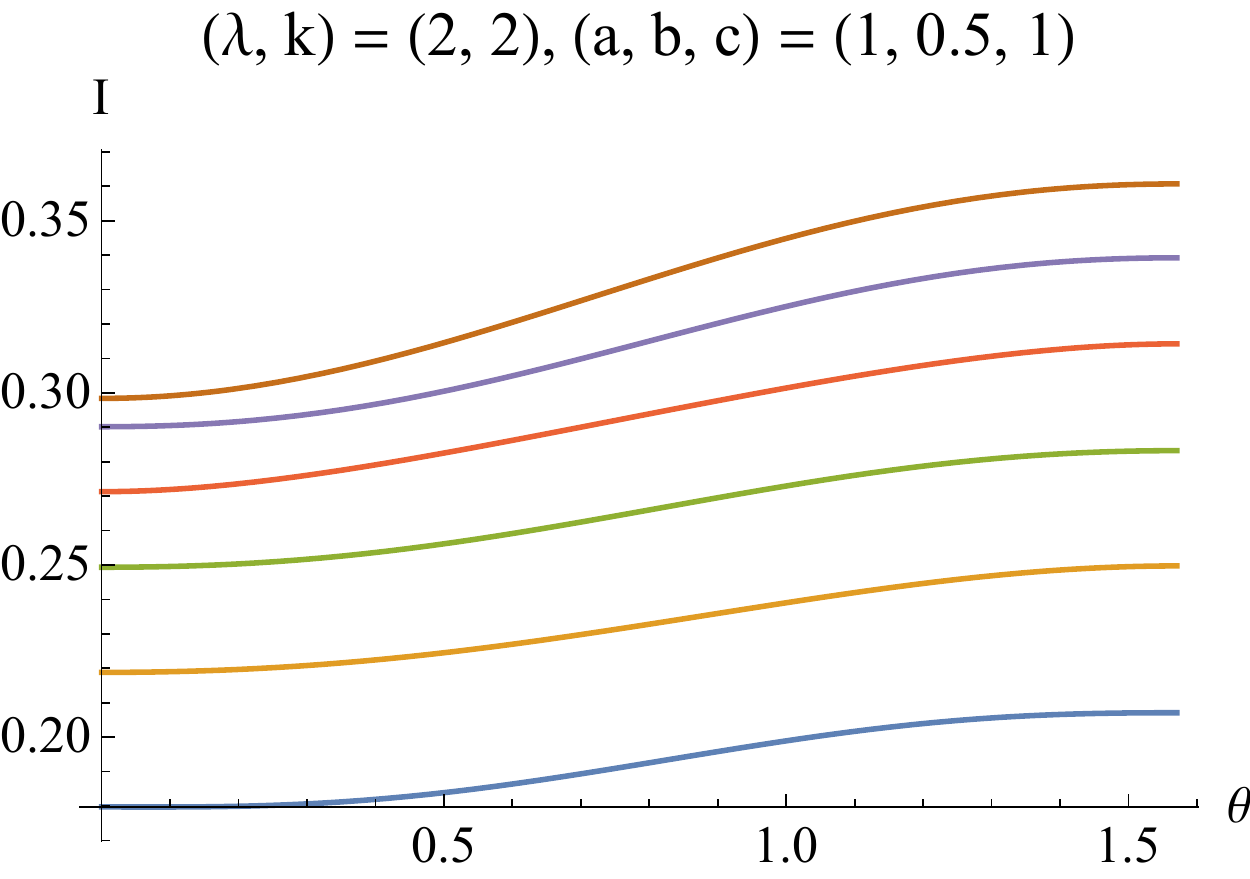}
	\includegraphics[width=0.49\textwidth]{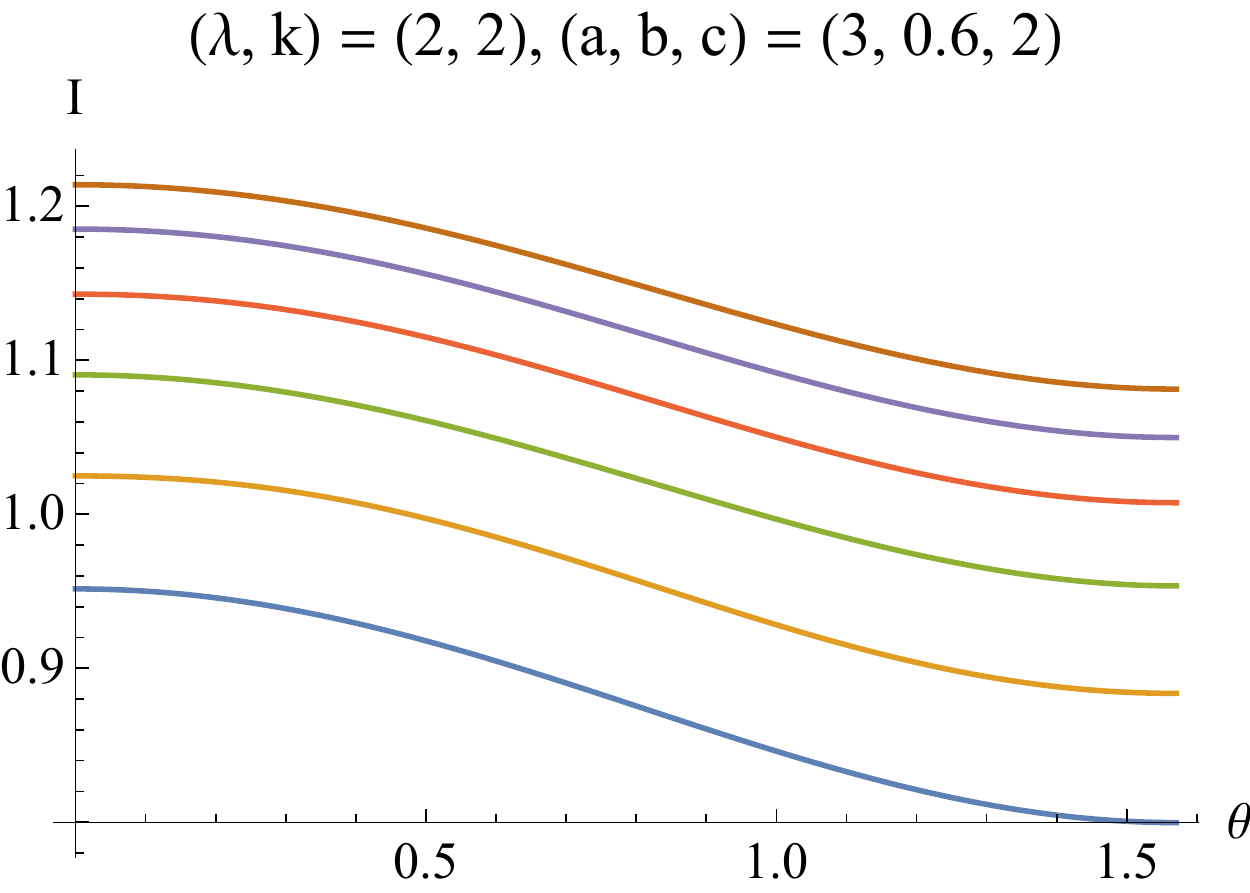}
	\includegraphics[width=0.95\textwidth]{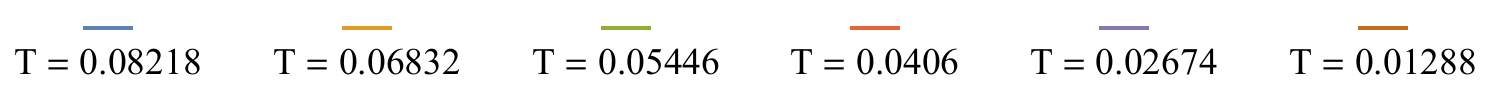}
	\caption{Each curve in above two plots is MI vs $\theta$ at $(\lambda,k) = (2,2)$ at temperature marked by the legends below. The left and the right picture corresponds to $(a,b,c) = (1,0.5,1)$ and $(3,0.6,2)$, respectively.}
	\label{fig:small_large_ell_vs_tem}
\end{figure}

The first interesting phenomenon we find is that the angular behavior of MI changes with the configuration. Fig. \ref{fig:small_large_ell_vs_tem} shows the angular dependence of MI for different configurations, from which we clearly see that MI increases with $\theta$ when the configuration is large; but for small configurations, MI decreases with $\theta$. Moreover, Fig. \ref{fig:small_large_ell_vs_tem} also illustrates that this monotonic phenomenon does not depend on temperature. Next, we examine a specific case $T=0.00892$ in detail to more clearly show the effect of configuration on the monotonicity of MI.

\begin{figure}[htbp]
    \centering
    \includegraphics[width=0.45\textwidth]{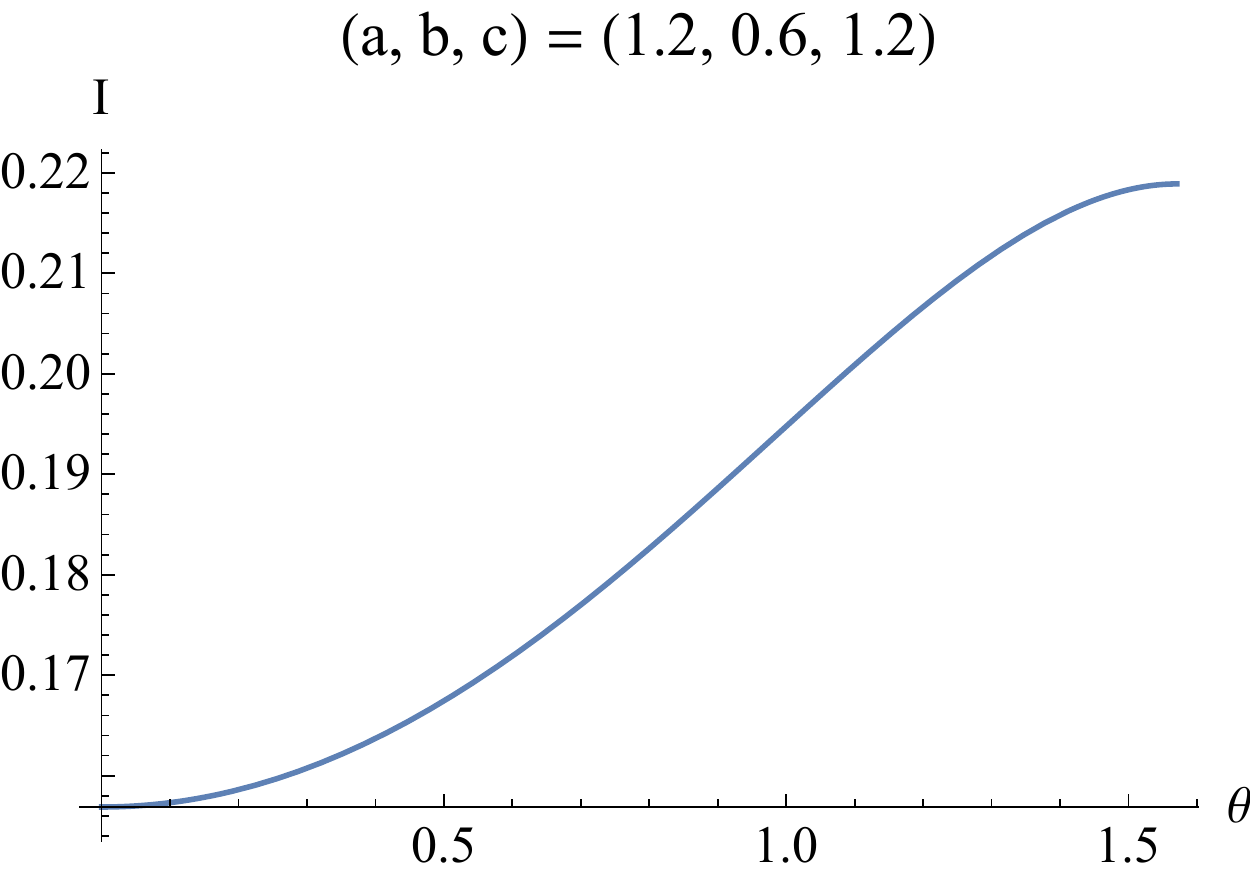}\quad
    \includegraphics[width=0.45\textwidth]{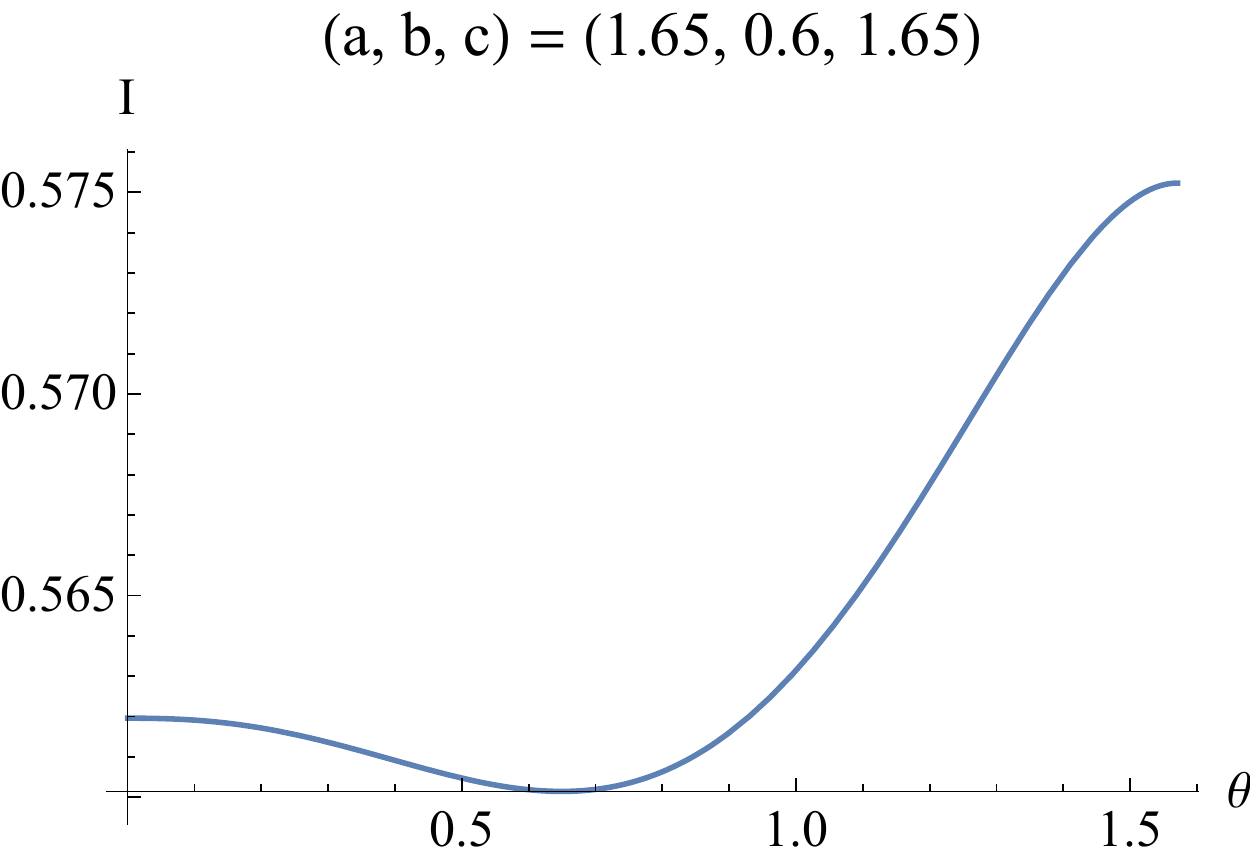}\\
    \includegraphics[width=0.45\textwidth]{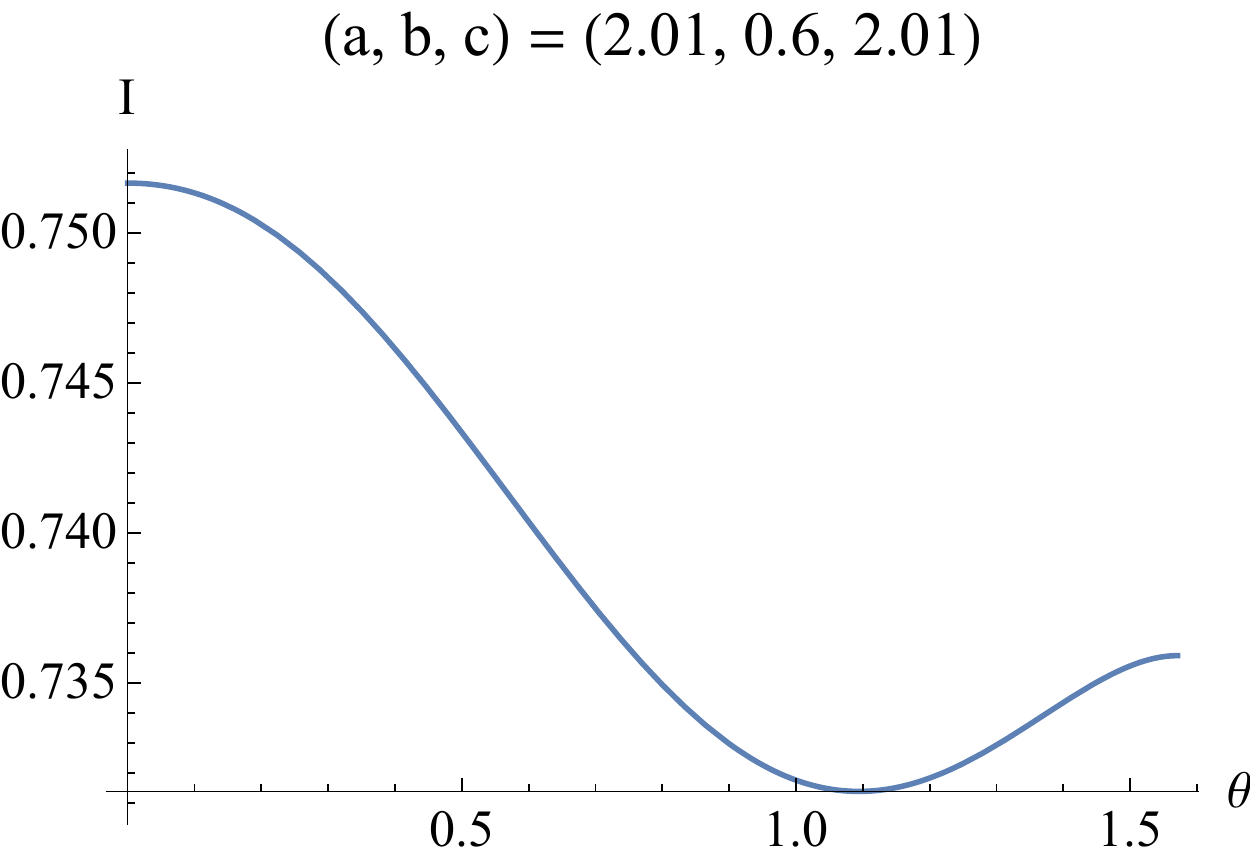}\quad
    \includegraphics[width=0.45\textwidth]{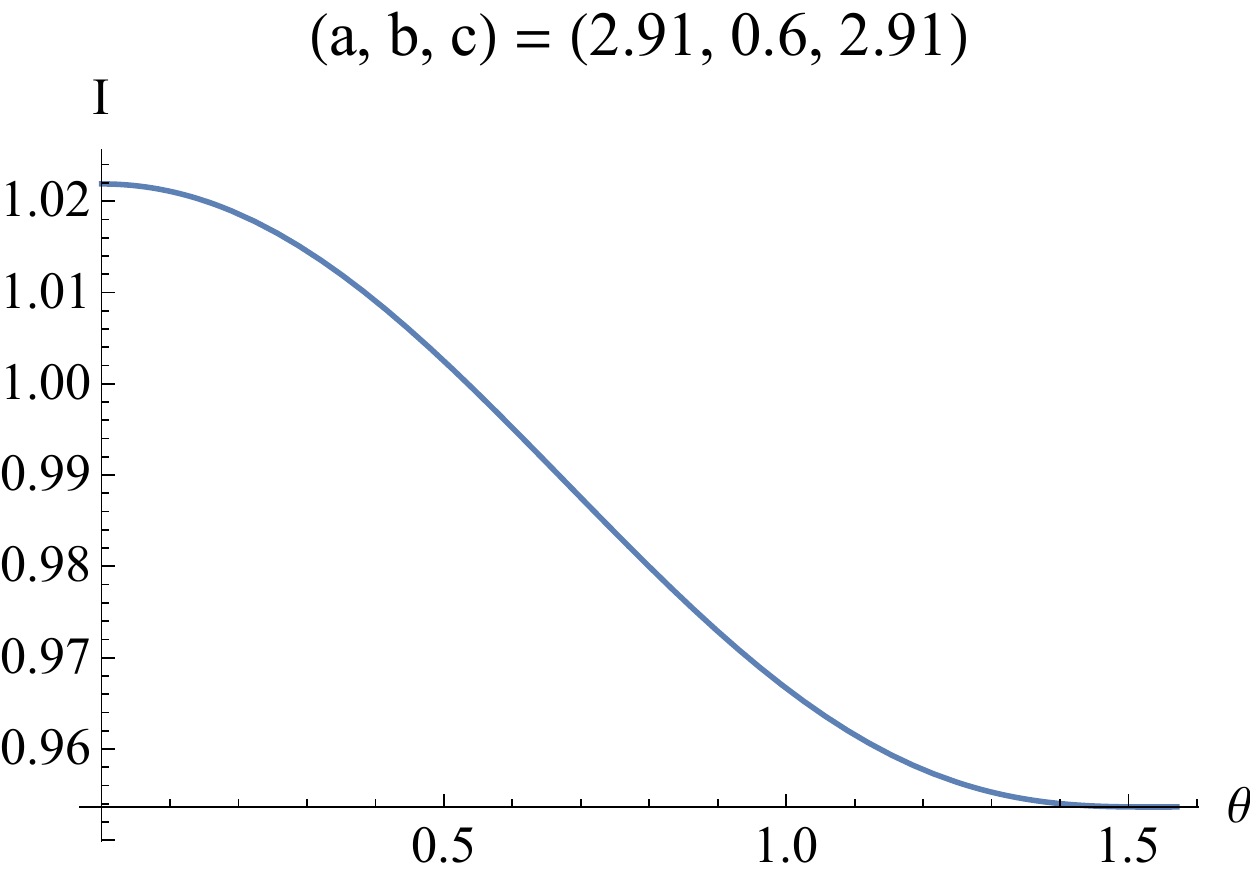}
    \caption{The plot is at $(\lambda,k) = (3,0.5)$ (insulating phase) and at $T=0.00892$, we also remark that the phenomena do not depend on temperature, therefore we demonstrate the phenomena at a specific value of $T$. We also remark that similar behaviors can be observed for metallic phases.}
    \label{fig:miwithellinsu}
\end{figure}

For simplicity, we set the lengths of $A$ and $C$ to equal. We demonstrate the phenomenon in Fig \ref{fig:miwithellinsu}, from which we see that the angular behavior of MI depends on the size of the subregion. At first, the MI increases with $\theta$ monotonously, which is contrary to the angular behavior of HEE. With the increase of $l$, however, the MI starts to become non-monotonic, and eventually monotonically decreases with $\theta$ at large enough subregions. In other words, the lattice suppresses the entanglement between small subregions; while for large subregions, the lattice enhances the entanglement.

The above size-dependence of the angular behavior $I(\theta)$ can be explained by the $\partial_{\theta} S$ at different ranges of $l$. The derivatives of MI with respect to $\theta$ reads,
\begin{equation}\label{theta-deriv}
\partial_{\theta} I= \partial_{\theta} S(A)+ \partial_{\theta} S(C) - \partial_{\theta} S(A\cup C).
\end{equation}
For small values of $a$ and $c$, $\partial_{\theta} S(A)$ and $\partial_{\theta} S(C)$ will be negligible because $S(A)$ and $S(C)$ are dictated by the asymoptotic AdS$_{4}$ region, and hence are insensitive to the anisotropy. Therefore, Eq. \eqref{theta-deriv} shows that the $\partial_{\theta} I$ is dominated by $-\partial_{\theta} S(A\cup C)$ for small values of $a$ and $c$, which explains the opposite angular behavior as that of the HEE. For large values of $a$ and $c$, however, the angular behavior will be the same as that of the HEE. The $S(A\cup C)$ is dominated by the near horizon geometry for large values of $a$ and $c$, \emph{i.e.} the thermal entropy density $\sqrt{g_{xx}g_{yy}}$. Therefore, $\partial_{\theta} S(A\cup C)$ is close to $0$ because $\partial_{\theta} \sqrt{g_{xx}g_{yy}}$ vanishes. Consequently, the angular behavior of MI will be the same as that of the HEE.

\begin{figure}[htbp]
	\centering
	\includegraphics[width=0.6\textwidth]{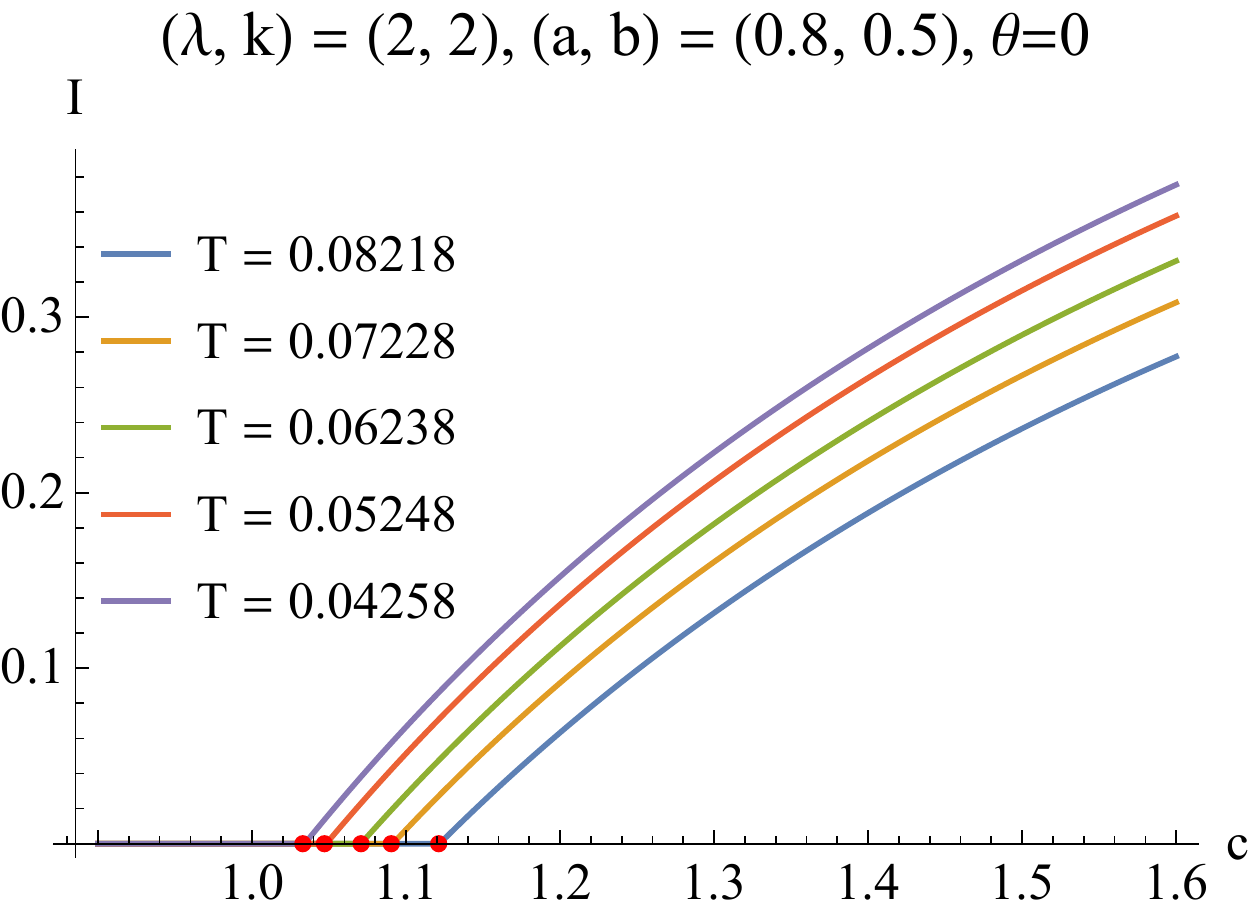}
	\caption{The critical values of $c$. The red dots are the location of $c_{c}$ for different temperatures.}
	\label{fig:showcc}
\end{figure}

\begin{figure}[htbp]
	\centering
	\includegraphics[width=0.45\textwidth]{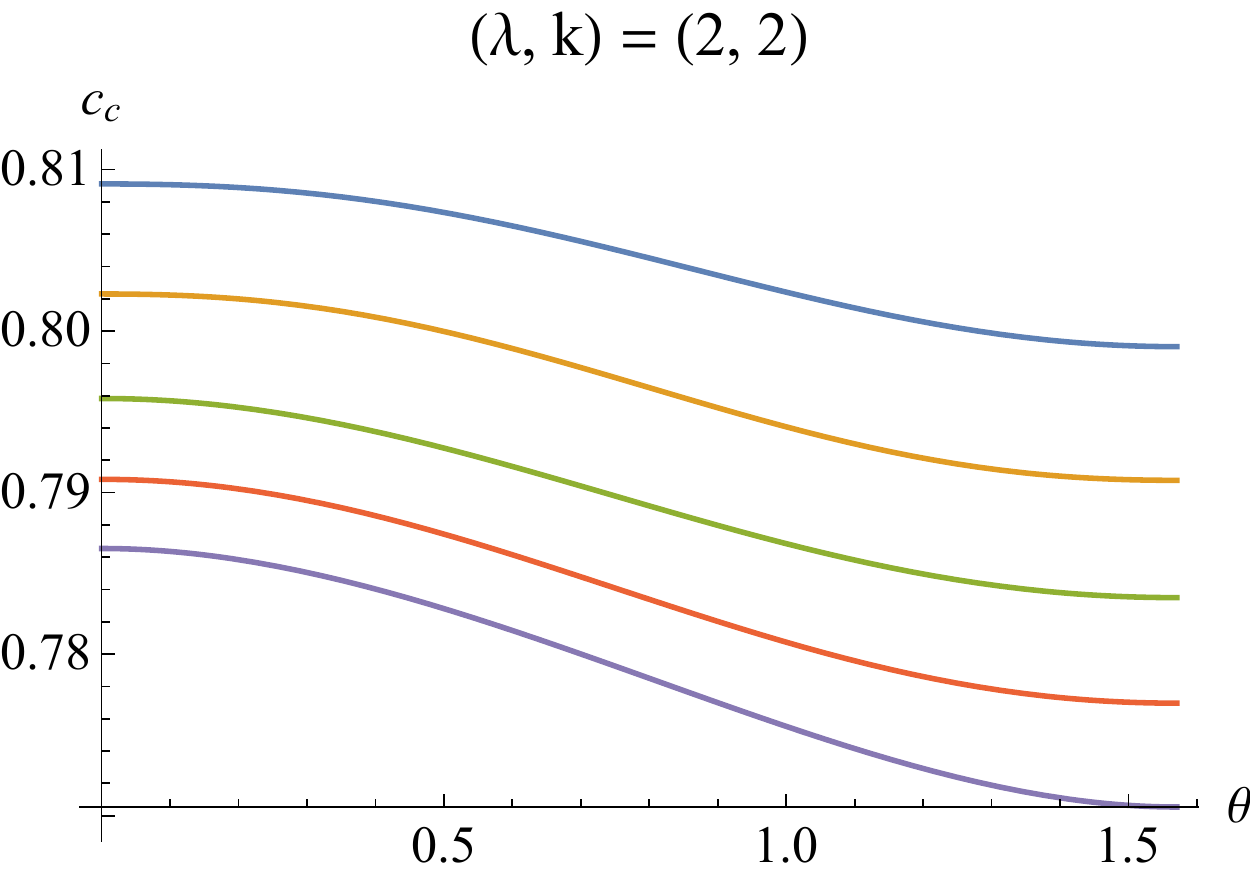}\quad
	\includegraphics[width=0.45\textwidth]{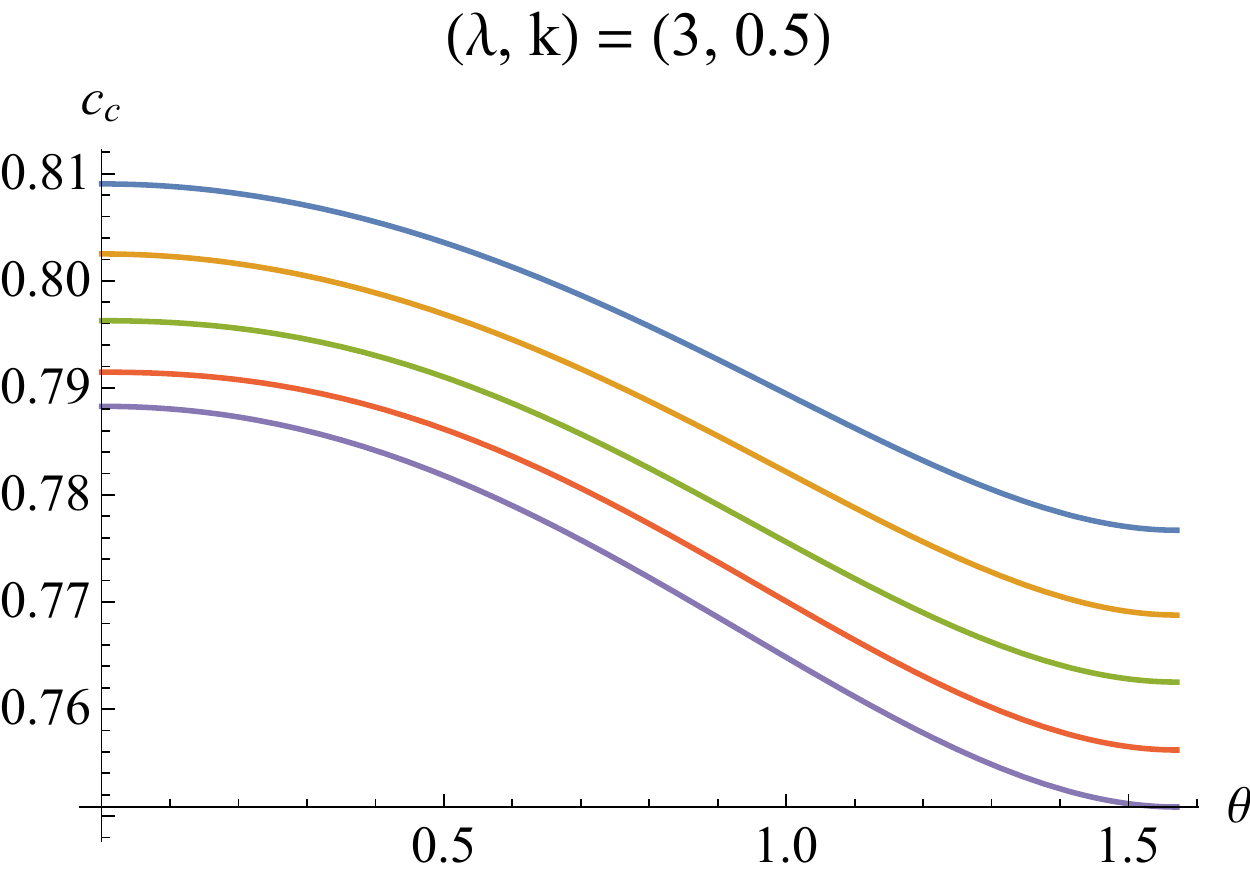}\\
	\includegraphics[width=0.8\textwidth]{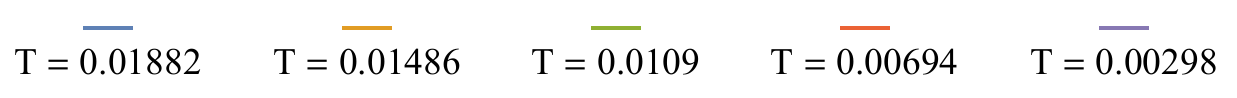}
	\caption{The $c_{c}$ for $(a,b)=(2,0.6)$.}
	\label{fig:ccformi}
\end{figure}

Another interesting quantity of the MI is the critical size of the subregion. Given $(a,b)$, there exists a critical value $c_{c}$ for the size of $C$. The MI is nontrivial only when $c>c_{c}$, otherwise MI vanishes (see Fig. \ref{fig:showcc} for a detailed demonstration of $c_{c}$). Next, we show the relationship between $c_{c}$ and $\theta$ in Fig. \ref{fig:ccformi}, We can see from the figure that $c_{c}$ monotonically decreases with angle, and this phenomenon has nothing to do with the temperature and the value of $(\lambda,k)$. Next, we argue that this phenomenon can be well-understood for small and large configuration limit.

For small values of $a$ and $c$, both $S(A)$ and $S(C)$ are insensitive to angle due to the asymptotic AdS$_{4}$ boundary. Therefore, the angular behavior of MI is mainly contributed by $-S(A\cup C)$. According to the monotonically decreasing beahvior of HEE, we conclude that MI increases with $\theta$. Therefore, subregion $C$ needs to downsize to maintain a nonzero MI. For large values of $a$ and $c$, however, $S(A)$, $S(C)$ and $S(A\cup B\cup C)$ are all insensitive to angle since the corresponding minimal surface are all approaching the horizon. Therefore, the MI increases with $\theta$, following from the increasing angular behavior of $-S(B)$ \footnote{By definition, $I(A,B,C) = S(A)+S(C)-S(B)-S(A\cup B\cup C)$. The angular behavior of $S(A)$, $S(C)$ and $S(A\cup B\cup C)$ are all trivial in large configuration limit, therefore the angular behavior of the MI is determined by that of $-S(B)$.}. Therefore, again, $c$ needs to decrease in order to maintain a nonzero MI. For intermediate values of $a$ and $c$, an analytical understanding of the monotonically decreasing behavior of $c_{c}$ still asks for further exploration.

\section{Discussion}\label{sec:discussion}

We studied the effect of anisotropy on HEE and MI in anisotropic Q-lattice model. We find that the lattice always enhances the HEE, which reflects how Q-lattice deforms the background geometry. We also find that the anisotropic HEE always characterizes the QPT. For MI we find the angular behavior is size-dependent, which can be understood from the angular behavior of HEE. Next, we point out several directions worth investigating further.

The first step to deepen our understanding of how anisotropy affects the HEE and the MI is to study more general anisotropic models.
For example, the axion models, Q-lattice models with two-dimensional lattice, scalar lattice models and ionic lattice models.
We also note that, however, the background can be both anisotropic and inhomogeneous for scalar lattice models and ionic lattice models. As a result, it could be much harder to study the entanglement measures.
Moreover, the anisotropic entanglement properties are tied to responses of quantum systems \cite{cai:2010,hannah:2012,gauger:2011}. Therefore we may study the DC conductivity, and figure out its connection to the anisotropic HEE and MI.

Besides studying more anisotropic models, the entanglement structure could be further explored by studying more entanglement measures.
For example, the entanglement of purification, complexity, negativity, R\`enyi entropy, and so on. We could expect these information-related quantities to reveal more novel characteristics of the anisotropic systems.

\section*{Acknowledgments}
Peng Liu would like to thank Yun-Ha Zha for her kind encouragement during this work. This work is supported by the Natural Science Foundation of China under Grant No. 11847055, 11805083, 11775036.

\end{document}